\documentclass[aps,prb,twocolumn,amsmath,amssymb,superscriptaddress,floatfix,bibnotes]{revtex4-2}

\usepackage[english]{babel}
\usepackage[dvipsnames]{xcolor}
\usepackage{graphicx}
\usepackage{hyperref}
\usepackage[utf8]{inputenc}
\usepackage[T1]{fontenc}
\usepackage{physics}
\usepackage{mathtools}
\usepackage{siunitx}
\usepackage{bm}
\usepackage{dcolumn}
\usepackage{appendix}
\usepackage{lipsum}
\usepackage{multirow}
\usepackage{placeins}
\usepackage{mhchem}
\usepackage{caption}
\usepackage{comment}
\usepackage{adjustbox}
\usepackage{pgfgantt}
\usetikzlibrary{shapes, arrows}

\newcommand{\ho}[1]{{\ce{H2O}#1}}
\newcommand{\so}[1]{{\ce{S2O}#1}}
\newcommand{\feho}[1]{{\ce{[Fe(H2O)_6]^{2+}}#1}}

\usepackage[normalem]{ulem}

\begin{document}
\title{Predicting The One-Particle Density Matrix With Machine Learning}
\author{S.~Hazra}
\affiliation{School of Physics and CRANN Institute, Trinity College, Dublin 2, Ireland}
\author{U.~Patil}
\affiliation{School of Physics and CRANN Institute, Trinity College, Dublin 2, Ireland}
\author{S.~Sanvito}
\email{Corresponding author: sanvitos@tcd.ie}
\affiliation{School of Physics and CRANN Institute, Trinity College, Dublin 2, Ireland}
\date{\today}

\begin{abstract}
Two of the most widely used electronic-structure theory methods, namely 
Hartree-Fock and Kohn-Sham density functional theory, both require the iterative 
solution of a set of Schr\"odinger-like equations. The speed of convergence of 
such process depends on the complexity of the system under investigation, 
{the self-consistent-field algorithm employed}
 and on the initial guess for the density matrix. An initial density 
matrix close to the ground-state one will effectively allow one to cut out many 
of the self-consistent steps necessary to achieve convergence. Here, we
predict the density matrix of Kohn-Sham density functional theory by constructing a 
neural network, which uses the atomic positions as only information. Such neural
network provides an initial guess for the density matrix far superior to any 
other recipes available. Furthermore,
the quality of such neural-network density matrix is good enough for the evaluation
of interatomic forces. This allows us to run accelerated {\it ab-initio} molecular
dynamics with little to no self-consistent steps.
\end{abstract}
\maketitle

\section{Introduction}
Density functional theory (DFT)~\cite{theorem,exchange,oxford} has been playing a 
central r\^ole in the electronic structure calculations of molecules and solids
for more than six decades. The DFT success boils down to the rigorous theoretical
framework~\cite{theorem}, to the availability of well-controlled approximations
of the exchange-correlation functional~\cite{JLadder}, to the multitude of numerical
implementations~\cite{VASP,QE,WIEN2K,Abinit,FHIAIMS,Siesta,PySCF}, and to the 
community rigor in benchmarking results~\cite{rigor,Lehtola2023}. In principle, for a given 
functional one can find the ground-state density and energy by functional minimization 
{with respect to the electron density,}
a procedure denoted as orbital-free DFT~\cite{condensed,orbital,recent}. However, the
lack of a universal and accurate approximation to a functional form of the 
non-interacting kinetic energy, a shortfall hardly mitigated by machine 
learning (ML)~\cite{KieronKE}, makes the widespread use of orbital-free DFT 
impractical. The problem can be circumvented by the Kohn-Sham (KS) construct~\cite{exchange},
in which the minimization of the functional is performed by solving an associated 
system of single-particle Schr\"odinger-like equations. The one-particle potential
entering the KS equations does, in turn, depend on the electron density. Hence, the 
solution is iterative and requires a multi-step cycle, where the electron density 
and the KS potential are continuously updated until converge is reached. Such 
self-consistent field (SCF) process is common to other electronic structure methods,
for instance the Hartree-Fock scheme, where one updates the coefficients of the
molecular orbitals~\cite{QuantumChem}. 

{In general the number of iterations required by the SCF process to achieve the desired accuracy depends on the 
system's complexity, the particular numerical DFT implementation, the SCF algorithm used and the initial trial electron density. 
Systems presenting a band gap are typically easier to converge than metals, since oscillations in the electron density during the 
iterative convergence are largely suppressed. Such oscillations can be damped by selecting appropriate ways to update the charge 
density from an iteration to the next, a process that in general largely determines the rate at which convergence is achieved. Note 
that, depending on the specific DFT numerical implementation, one may decide to mix the KS Hamiltonian instead of the charge 
density. In any case, regardless of the quantity chosen for the SCF algorithm, typically the output of several 
previous iterations are combined to determine the new input. The direct inversion of the iterative subspace (DIIS) method, proposed 
by Pulay \cite{diis1,diis2}, is probably the most widely used SCF-solver in modern local-orbital DFT codes, where one mixes the Fock
matrices. However, there exists a multitude of alternative schemes and refinements, whose performance depends on
the specific problem at hand 
\cite{blackboxscf,damping,outperformdiis,levelshift,ediis,Bhattacharyya,RoothaanHallenergyfunction,Coiterative,SOSCF}.}

Regardless of the mixing method used, an initial guess for the charge density
close to the final self-consistent solution will speed up convergence, by typically 
reducing the number of iterations to perform. Such an initial density is usually 
defined either over a real-space or a $k$-space grid, in DFT codes based on 
plane waves, or in the form of a one-particle density matrix (DM), for local-orbital codes
\cite{Lehtola1}. There are multiple strategies to generate 
the initial charge density (or the DM), which all reduce to solve an
associated non-self-consistent problem of some kind. As several of such schemes
will be explored here, a detailed description will be provided in our method 
section. 

The main aim of our work is to construct ML models, namely
neural networks, to learn the ground-state one-particle DM of a DFT calculation.
The models are based on structural and atomic information only, namely the 
chemical nature and positions of the atoms forming a molecule. The so-constructed 
DM can then be used either as a starting point for a SCF cycle or, if the accuracy 
is good enough, to perform a non-self-consistent evaluation of the various
observables, for instance energy and forces. 

Note that ML schemes to generate 
the charge density in either real space~\cite{BurkeD,Chandrasekaran,Ellis,Bruno} 
or over an atom-center basis~\cite{Grisafi} have been already proposed. One can 
then, in principle, take one of such models and try to construct the DM from the 
computed charge density. This strategy, however, requires projection across different 
incomplete basis sets, a process that inevitably introduces additional errors.
These are likely to be large enough to preclude the use of the ML
DM, namely it will be unlikely competitive with other initial-density generation 
approaches. Furthermore, learning the DM
directly enables the straightforward evaluation of the expectation values of all 
one-particle operators. These include non-local potentials, so that the same scheme
can be used with Hartree-Fock calculations. Note that a mapping between the 
external potential and the one-particle density matrix, constructed over a 
kernel ridge regression, has been recently proposed ~\cite{Pavanello}. 
This is complementary to our work, since it requires smaller training
sets for similar-sized molecules, but the inference is significantly more demanding.
Alternatively, there is a body of literature looking at the construction of
the Hamiltonian, most typically the Kohn-Sham one, from an equivariant description of
the molecule's structure. This is typically achieved either through an equivariant network
or data augmentation. In both cases, the result is that of having rather large and deep models,
which need to be trained over extensive data sets~\cite{Schutt2019,Unke2021,Zhang2022}.

The paper is organised as follows. In the next section we discuss the main 
methodological aspects of the models construction and the DFT implementation
used to generate the data and benchmark the results. Then, we proceed by presenting
the results. In particular, we first evaluate the quality of our DM 
as starting point for a SCF cycle, and then we analyse the error on the predicted
energy and forces. With these results at hand we perform non-self-consistent 
{\it ab initio} molecular dynamics, whose results are compared with fully
converged one. Finally, we conclude and suggest possible future directions.

\section{Methods}

{As discussed in the introduction, our task is to predict the converged DM of a DFT calculation by 
using a ML model, which utilises only chemical and structural information of a molecule. In particular, we consider
fully-connected dense neural networks together with global structural descriptors, and we predict all the independent 
matrix elements of the DM. The models are defined by the neural-network architecture, the descriptor types and
the content of the training and test sets, all aspects that will be described here in detail.}

\begin{figure} [!ht]
    \centering
    \includegraphics[width=3.5cm,height=3.2cm]{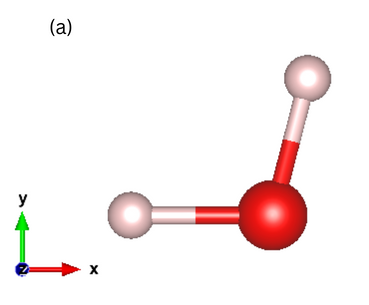}
    \includegraphics[width=3.5cm,height=3.2cm]{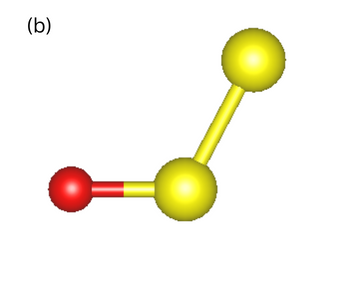}
    \includegraphics[width=3.5cm,height=3.2cm]{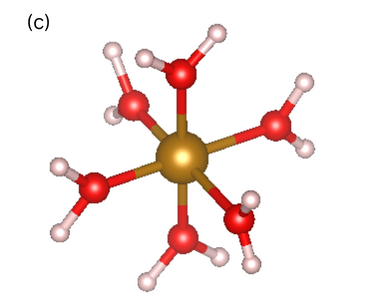}
    \caption{The molecules investigated in this study placed in their default positions along
    the Cartesian axes: (a) \ce{H2O}, (b) \ce{S2O}, and (c) \ce{[Fe(H2O)_6]^{2+}}. 
    Color code:  H = white, O = red, Fe = dark golden, S = yellow, $x$-axis = red, $y$-axis = green, 
    $z$-axis = blue.}   
    \label{fig:plot_1}
\end{figure}

All the electronic structure theory calculations performed in this work
to generate the training dataset and to benchmark the results have been 
produced with the open source Python package, PySCF \cite{software,PySCF}.
{PySCF implements all-electron DFT and a number of quantum-chemistry methods, such as Hartree-Fock, 
over a Gaussian basis set.}
For all our calculations 
we have used the cc-pVDZ basis~\cite{Dunning1989}, formed by double-zeta polarised orbitals for 
the valence electrons. Thus, for the atomic species relevant for our tests
we have 5 basis functions for H, 14 for O, 18 for S and 43 for Fe.
In fact, three different molecule have been considered in this work, namely 
\ho, \so\ and \feho, see Fig.~\ref{fig:plot_1}. 

We start our analysis with \ho\ ($C_{2v}$ symmetry), 
since this presents a simple electronic structure and typically no convergence 
problems. Its charge density, however, does not differ much from a superposition
of atomic densities, so that a more stringent test is provided by \so\ 
(also $C_{2v}$ symmetry). Importantly, both \ho\ and \so\ can be described by 
only three structural features, so that the more complex \feho\ ($O_{h}$ symmetry) 
is investigated last. The \feho\ molecule also gives us the opportunity to test 
our method for a metal bond.
All electronic structure calculations are performed with the BLYP functional, which 
combines the generalised gradient approximation for the exchange energy of Becke~\cite{B88} 
with the Lee-Yang-Parr correlation energy \cite{LYP}, 
{as implemented in the {\sc libxc} library~\cite{LibXC}.} 
{When creating the training, validation and test sets, the SCF cycle is converged with the DIIS scheme~\cite{diis1,diis2} for 
\ho\ and \so, while for \feho we employ a second-order solver (SOS)~\cite{Coiterative,SOSCF}, as the convergence appears more problematic.
In contrast, when analyzing the convergence history of different initial DM guesses, we will consider both the DIIS and SOS SCF algorithm.}

{Since our ML DM will be compared with that generated by conventional initial guesses, it is worth spending some time in describing these.
Possibly, the simplest choice constructs the charge density as a superposition of atomic densities, while the DM is obtained from the orbitals 
that digonalize the Fock matrix associated to such spin-restricted guess density. This is the `minao' PySCF default option~\cite{Almlof1982,Lenthe2006}.
Alternatively, one can use the eigenstates of the non-interacting problem, namely those orbitals that diagonalize an Hamiltonian comprising only the
kinetic energy and the nuclear potential. This is the one-electron DM, `1e' option in PySCF, which usually represents a poor starting point for 
molecules~\cite{Lehtola1}. Then, there are options based on spin-restricted atomic Hartree-Fock calculations, employing different recipes for the 
construction of the guess orbitals used to compute the DM. In PySCF these are called `atom' \cite{Lehtola2} and `huckel' \cite{Lehtola1}. 
Finally, one can construct the DM with the orbitals obtained from the solution of a superposition of tabulated atomic potentials~\cite{Lehtola2}.
This is the `vsap' option~\cite{Lehtola1}.}

\begin{table*}[!t]
\caption{Table summarising the structure and performance of the final neural networks trained 
for the three molecules. Here we report the number of features defining the input, $N_\mathrm{in}$, the
dimension of the DM, $D_\mathrm{DM}$, the number of the network hidden layers, $N_\mathrm{hi}$, the total
number of neurons forming each hidden layer, $N_\mathrm{nu}$, and the total number of weights, $N_\mathrm{w}$.
Then, we report the mean absolute error (MAE), the largest error on the matrix elements,  $\delta \rho_\mathrm{max}$, 
the root-mean square error (RMSE), and the $R^2$ coefficient of the DMs. All errors refer to the test sets and 
they are in atomic units (a.u).} 
\label{table1}
\begin{ruledtabular}
\begin{tabular}{cccccccccccc}
 Molecule& $N_\mathrm{in}$ & $D_\mathrm{DM}$ & $N_\mathrm{hi}$ & $N_\mathrm{nu}$ &($N_\mathrm{w}$) & MAE& $\delta \rho_\mathrm{max}$ &RMSE & $R^{2}$ \\ \hline
 \ho~&$3$ &$24\times24$&$2$& $18, 32$ &662 & 0.0002&0.0057&0.0003& 0.9999 \\
 \ce{S2O}&$3$ &$50\times50$&$2$& $18, 28$ & 583 &0.0002& 0.0112& 0.0003& 0.9999\\ 
 \ce{[Fe(H2O)_6]^{2+}}&$6$ &$187\times187$&$2$ & $16, 32$ & 640& 0.0002 & 0.0283 & 0.0005& 0.9993 \\
\end{tabular}
\end{ruledtabular}
\end{table*}

Here we predict the initial guess DM with dense neural networks, where the input features are 
the independent Cartesian coordinates of the molecules. A summary of the structure of the different 
neural networks, together with the training-set errors, are reported in Table~\ref{table1}. 
The use of the Cartesian coordinates together with a dense neural network effectively forces
an equivariant quantity, namely the DM, to be described by an invariant model. This issue
is here resolved by manually removing the rotational and translational degrees of freedom of 
the molecule, a procedure that makes the entire problem invariant. Of course, such solution
is not general and a more elegant way to tackle the problem would be to use a fully 
equivariant representation of the molecular structure~\cite{Tess}. This, however, adds a 
significant new layer of complexity, which we wish to avoid for our early study.
Thus, we remove the translational degrees of freedom by fixing a given atom at the
origin. In particular, we use oxygen, the central sulphur and the \ce{Fe^{2+}}
cation, respectively for \ho, \so\ and \feho. Then the rotational degrees of freedom
are imposed by selecting an appropriate rotation. For the triatomic molecules, \ho\ and
\so, we constrain one atom on the negative $x$-axis and the second one in the $x$-$y$ 
plane, so that the molecules are described by only three coordinates. In contrast, 
for \feho\ we force the O atoms of the \ho\ ligands to be on the three Cartesian axes
and we consider only variation in the Fe-O bond length (the water molecules are taken
as rigid). This returns us six independent coordinates (see Fig.~\ref{fig:plot_1}). 

The structure of the neural network has been optimized by varying the number of hidden 
layers and their size, by minimizing the mean absolute error (MAE). The optimal configuration
for each molecule is reported in Table~\ref{table1}. Note that in all cases we employ the exponential 
linear unit activation function. The datasets used to construct the model are formed by 9,000
configurations for training, 800 for validation and 1,000 for testing. In the case
of \ho\ and \so\ such configurations are extracted from {\it ab-initio} Born-Oppenheimer molecular 
dynamics trajectories at 150~K, also performed with PySCF. In particular, we run for 117 and 130 
picoseconds, respectively for \ho\ and \so, by using the Nose-Hover thermostat through the pyLammps API
as implemented in the LAMMPS package~\cite{lammps}. In contrast, for the case of \feho\ we consider 
random rigid displacement of the \ho\ molecules, such that the Fe-O bond length is varied within 10\% 
from its equilibrium value (2.0525~\AA). These return us training-set mean absolute errors (MAEs) of 
the order of $10^{-3}$ atomic units (a.u.). Note that typically, the largest DM matrix elements are found 
along the diagonal and they can reach values close to unity, while a significant fraction of the off-diagonal 
matrix elements remain small. For example, for the \ho\ molecule we find 8.15\% of the matrix elements, $\rho$, 
having values  $0.1<|\rho|<1$, 43.75\% in the range $10^{-3}<|\rho|<0.1$ and 48.1\% being $|\rho|<10^{-3}$.
The parity plots associated to our neural networks, together with the typical matrix 
elements distributions can be found in the Appendix for all the three molecules. 
Note also that our numerical construction of the DM does not guarantee idempotency to be 
satisfied, and in fact, we find that this is numerically violated. Since idempotency is a non-linear condition 
it is difficult to implement it as a constrain in the neural networks. Such drawback is here compensated by
the numerical accuracy achieved, as we will demonstrate in the following.

Finally, we perform tests on how the computed DM can drive molecular dynamics, namely 
we perform {\it ab-initio} molecular dynamics using our ML DM and not the one
resulting from a SCF cycle. In this case special care must be taken, since the neural
networks return DM only for molecules having the specific spatial orientations described
before. For this reason, we operate the following workflow. A molecule at an arbitrary 
position is translated back to the origin and rotated so to have the orientation required 
by the neural networks. Then, the DM is evaluated with the network and used as a starting 
guess for a static DFT calculation (non-self-consistent). Energy and forces are thus evaluated 
using PySCF. The molecule is then rotated/translated back to the original position and the
same rotation is applied to the forces. Such force field is input into the molecular dynamics
package, which updates the atomic coordinates. Then the process is repeated. The molecular
dynamics steps are implemented in the LAMMPS package \cite{lammps}.
{Although more cumbersome than standard MD, the strategy adopted here allows us to use our simple structural descriptors for a problem, 
the construction of the DM, which is intrinsically translational invariant and rotational covariant. Translational invariance can be achieved by using local 
structural descriptors, while rotations can be accounted for with a covariant model. Here we have preferred to keep our model as simple as possible 
so to concentrate fully on demonstrating how a DM can be constructed with ML.}

\section{Results and Discussions}
In order to validate our entire approach we perform three different tests. Firstly, we 
evaluate the efficacy of the ML DM as a starting point for a SCF cycle.
Then, we quantify the accuracy of the DM in determining energy and forces. Finally,
we compare the molecular-dynamics trajectories driven by the forces associated to the 
ML DM with those of fully self-consistent {\it ab-initio} molecular 
dynamics.

\subsection{Machine-learning DM as initial guess}
\begin{figure}[!ht]
\includegraphics[width=\columnwidth]{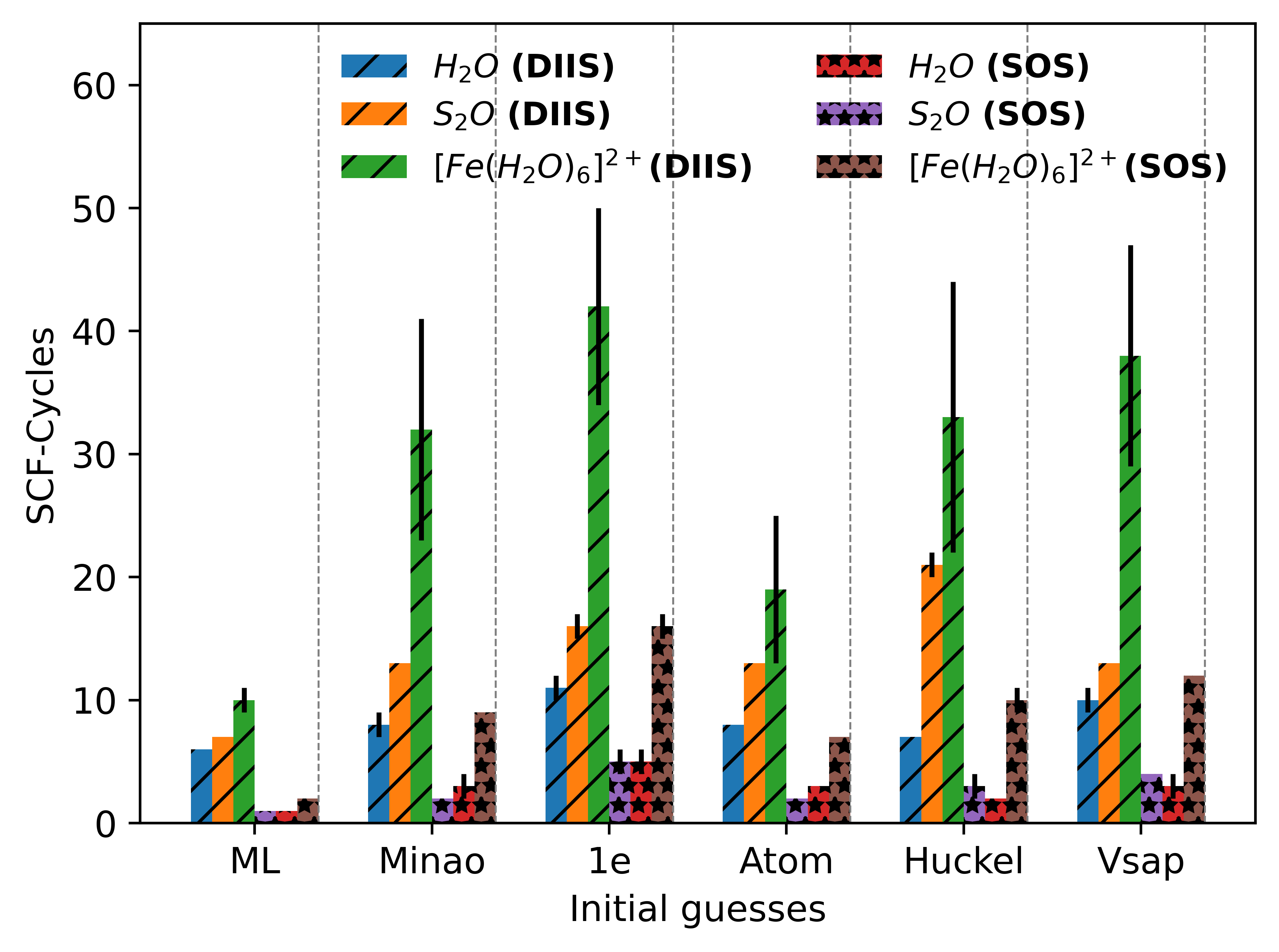}\\
\caption{Total average number of SCF steps taken to achieve convergence for 
different starting DMs and mixing schemes. Here the convergence criterion is on the 
total energy between two consecutive SCF steps that should be lower than $10^{-9}$~Ha.
The black bars around the mean indicate the variance. Variances lower than
one iteration are not displayed.}  
\label{SCF_accelaretion}
\end{figure}

The first test consists in evaluating how accurate is the DM generated by the neural 
networks as starting point for the DFT SCF cycle. The test is performed over 1000
new configurations for each molecule and in Fig.~\ref{SCF_accelaretion} we report the average
number of SCF iterations performed to achieve convergence and their associated variance. 
In this case converge is defined by having an energy difference between subsequent 
iterations lower than $10^{-9}$~Ha, a value that sets a rather tight convergence criterion. 
For this test we perform two sets of calculations, where the SCF cycles are driven respectively 
by the DIIS or the SOS mixing scheme, with convergence parameters set by the PySCF default. 

In general we find the SOS DM-update strategy to be significantly more performing than the
simpler DIIS, with the total number of iterations reducing by approximately a factor
three regardless of the molecule or the initial DM. Note that this advantage is partially 
compensated by the SOS scheme being numerically heavier than DIIS, namely a single iteration 
takes longer. We have also found that sometimes for \feho\ and the DIIS solver, 50 iterations 
are not enough to achieve convergence. This is somehow expected considering the electronic
structure of the \feho\ cation. In fact, \feho\ is a spin-crossover molecule presenting a 
temperature-induced low-spin to high-spin transition, driven by a distortion of the octahedral 
coordination shell of the Fe$^{2+}$ ion. This is only partially described by DFT~\cite{Andrea},
and a multi-determinant theory appears more appropriate~\cite{CIPaper}. For this reason it
is not surprising that for some highly distorted configurations our non-spin-polarised DFT 
calculations struggle to converge. Note that this is not so crucial here, since we are not
seeking to compute the exact ground state of \feho, but simply to present a test example for
`difficult' convergence. In any case, when convergence is not achieved, the SCF cycle is stopped 
after 50 iterations. 

Despite these differences, the convergence-speed trends with respect to the initial DM are rather 
similar across the two mixing schemes, so that in our discussion we refer first to data 
obtained with the SOS algorithm. As expected the \ho\ and \so\ molecules converge significantly faster 
than \feho, as their simple covalent bonding structure would suggest. Also as expected, the simple
`1e' default provides the worst starting DM and convergence is achieved in five iterations for
\ho\ and \so\ and about 16 for \feho. The other conventionally constructed starting DMs appear
to perform rather similarly to each other with the two covalently bonded molecules converging in
about 3-4 SCF steps, and \feho\ in about 10. Most importantly, our DM significantly outperforms 
any other methods, with a single SCF iteration being necessary for \ho\ and \so, while \feho\ 
requires only two. This gives us a speedup in the computation of the SCF cycle comprised between
a factor 3 and a factor 5. Note that the speedup is less pronounced when the DIIS mixing scheme
is used, in particular for the covalently bonded molecules, where the advantage over the other
schemes is only fractional. This difference seems to boil down to the inefficiency of the mixing scheme, 
which brings the iteration count to 6-7 even when the calculation is initiated with the ML DM. 
{While it is not straightforward to establish why the SOS algorithm offers a better convergence speedup to ML-generated initial DMs',
we note here that, in general, DIIS algorithms may not honour well the initial guess, meaning that the optimisation procedure may lead the electron 
density anywhere in the variational space. It is then expected that such drawback penalises more DM's close to the fully converged one than more inaccurate 
ones.}

\subsection{Convergence analysis}

We now look in more detail at how convergence is achieved for different starting DMs and the two
different mixing schemes. We begin by considering \ho\ and then move to \feho. The results
for \so\ are somehow in between these two cases and {are presented in the Supplementary 
Information (SI).} 
In figure~\ref{h2o_ener_den_scf}, we show the
total energy (with respect to the ground-state energy) as a function of the iteration number, $n$, for 
\ho\, computed with the DIIS [panel (a)] and SOS [panel (b)] mixing scheme. Furthermore, in panel (c) 
we present the norm of the difference between the ground state (converged) DM, $\rho^\mathrm{GS}$, 
and that at the $n$-th iteration, $\rho^n$, also along the DIIS-driven SFC cycle. This last quantity is computed 
as $\Delta\rho=\sum_{ij}|\rho^\mathrm{GS}_{ij}-\rho^n_{ij}|$, with $\rho_{ij}$ being the $i, j$ DM matrix element. 
Although the detail of each SCF cycle may differ depending on the molecule geometry, the figure shows 
a typical case.
\begin{figure*} [!ht] 
    \centerline{\includegraphics[width=0.30\textwidth]{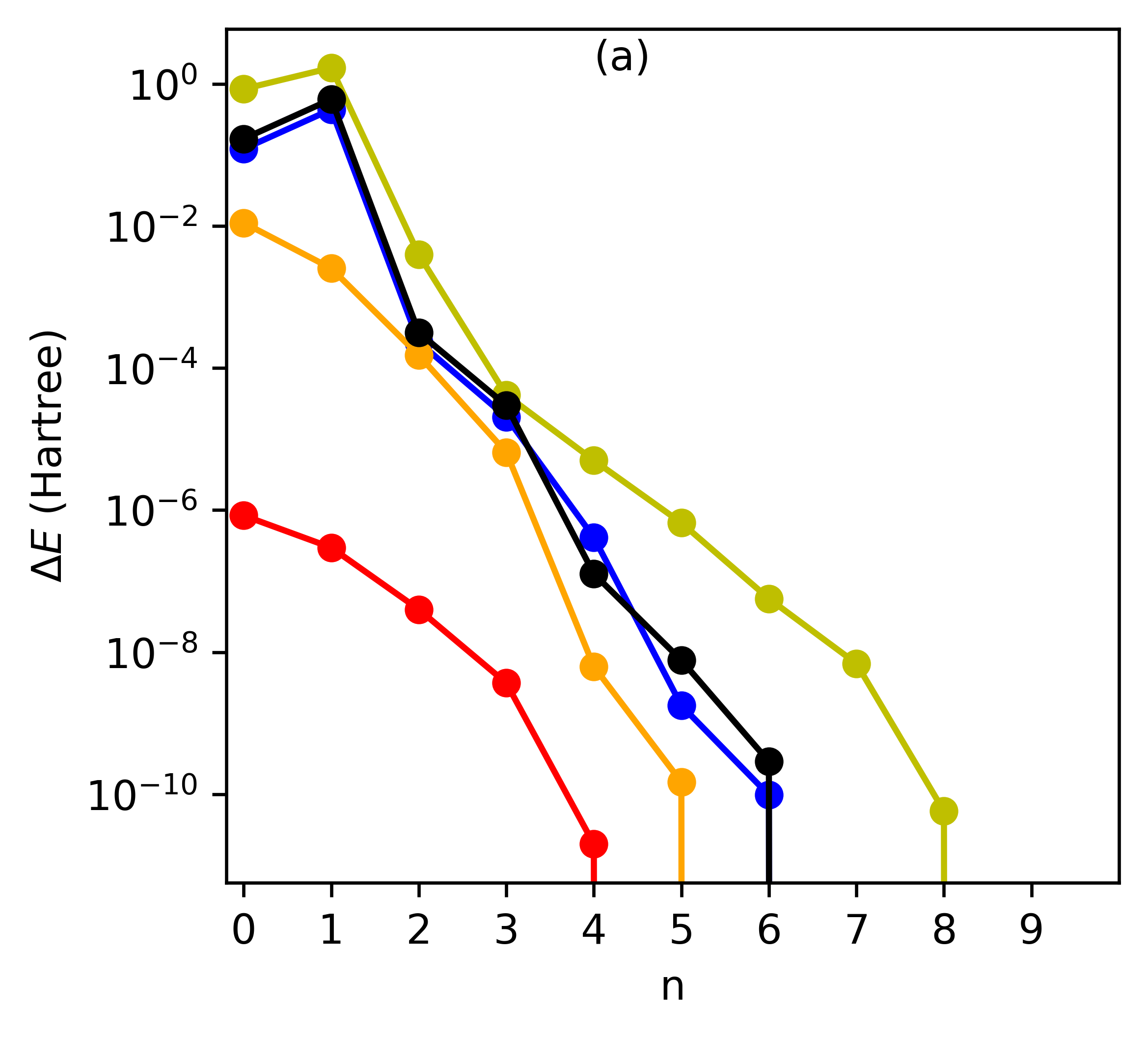}
    \includegraphics[width=0.30\textwidth]{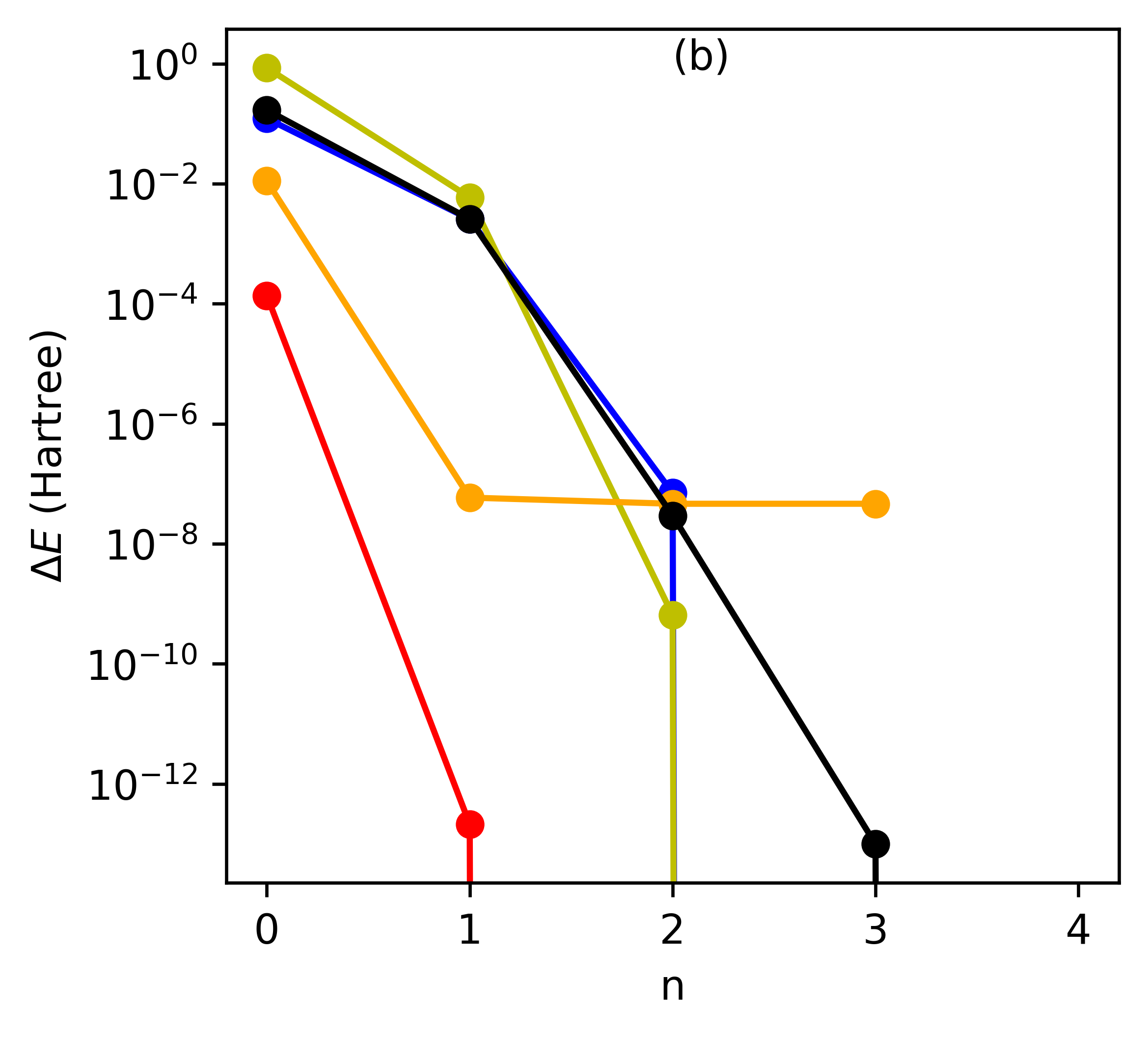}
    \includegraphics[width=0.30\textwidth]{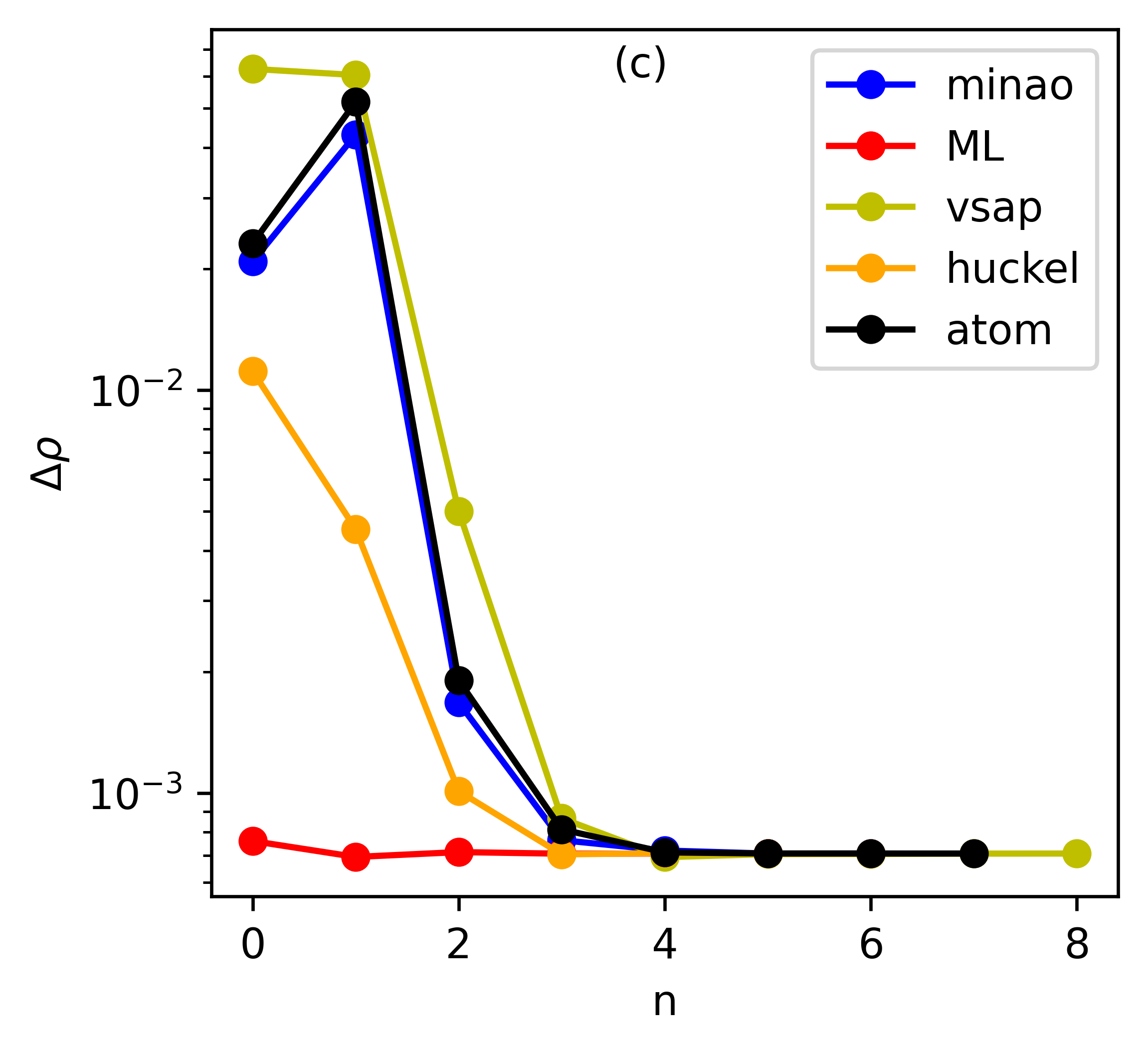}}
    \caption{Analysis of the SCF cycle for \ho. In panels (a) and (b) we show the total
    energy (measured with respect to the ground-state energy) as a function of the iteration number, 
    $n$, for convergence driven by the DIIS and SOS mixing scheme, respectively. In panel (c) we 
    present the norm of the difference between the ground-state converged DM, $\rho^\mathrm{GS}$, 
    and that computed at the $n$-th iteration, $\rho^n$. In this case we follow the DIIS-driven SCF cycle. 
    For ease of visualization in all plots the $y$ axis is on a logarithmic scale.}  
    \label{h2o_ener_den_scf}
\end{figure*}

In general all initial DMs are somewhat different from the final ground-state one, with the largest
variations found, as expected, for the `1e' initialization (the total energy difference at $n=0$ is
in excess of 8~Ha and it is not displayed here). The convergence is then monotonic,
when the SOS solver is used, while it may present oscillations for DIIS. This explains the
largest number of iterations typically taken by DIIS. Strikingly the ML-generated DM appears
extremely close to the final ground-state one, so that the convergence is monotonic in all 
cases. In fact, the $n=0$ computed total energies for \ho\ are on average within $10^{-4}$~Ha
from their ground-state value and the percentage variation of the DM at the first iteration is only 
0.196\%. This suggests that, by large, the ML-DM already provides an excellent 
estimation of the ground-state density matrix. As a comparison the second best initial
DM appears to be that generated with a restricted Hartree-Fock calculation (`huckel' option), 
with an initial total energy error of about $10^{-2}$~Ha. All the others DM-generating schemes 
have an initial error larger than 0.1~Ha.

\begin{figure*}
    \centerline{\includegraphics[width=0.30\textwidth]{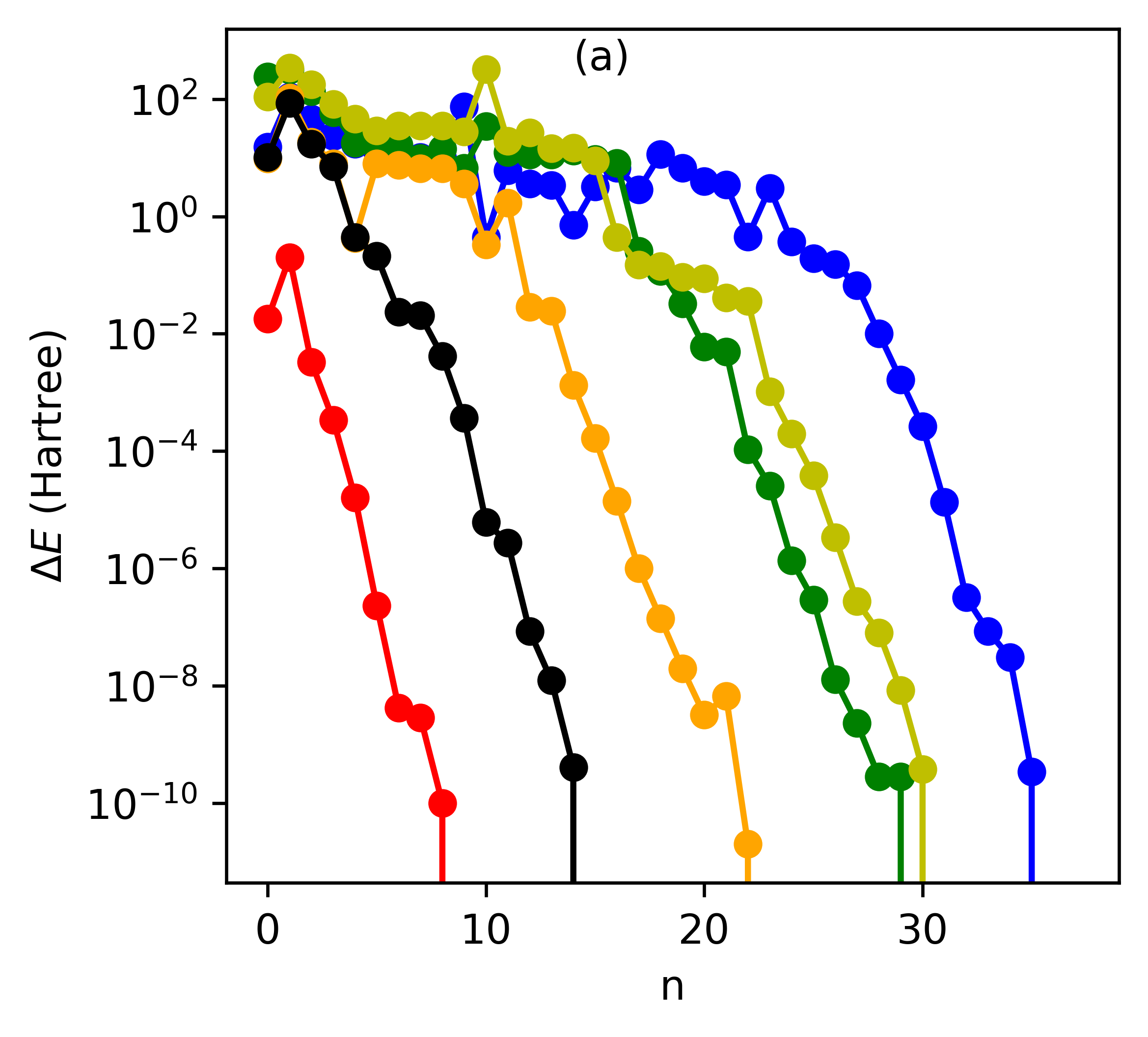}
    \includegraphics[width=0.30\textwidth]{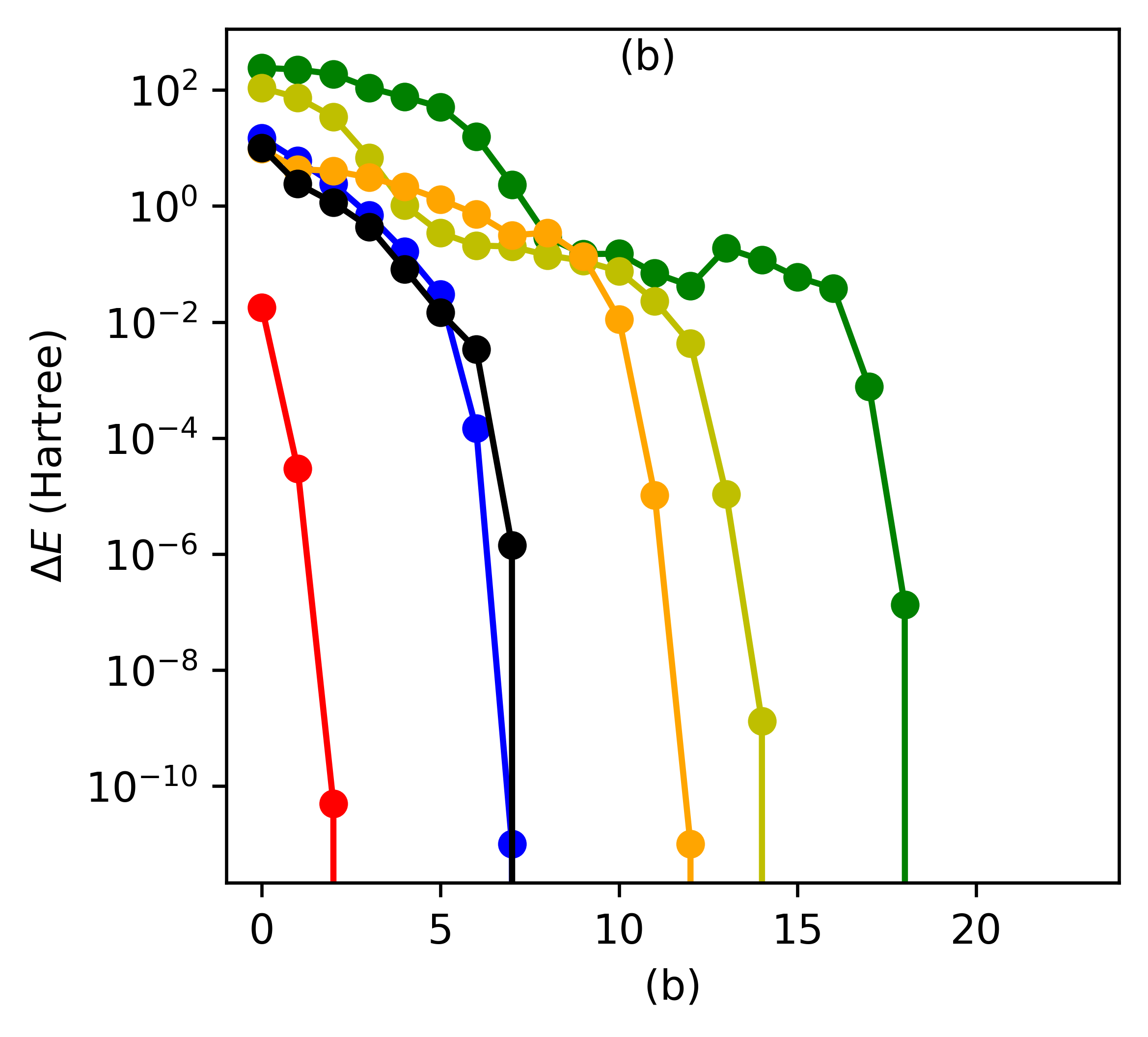}
     \includegraphics[width=0.30\textwidth]{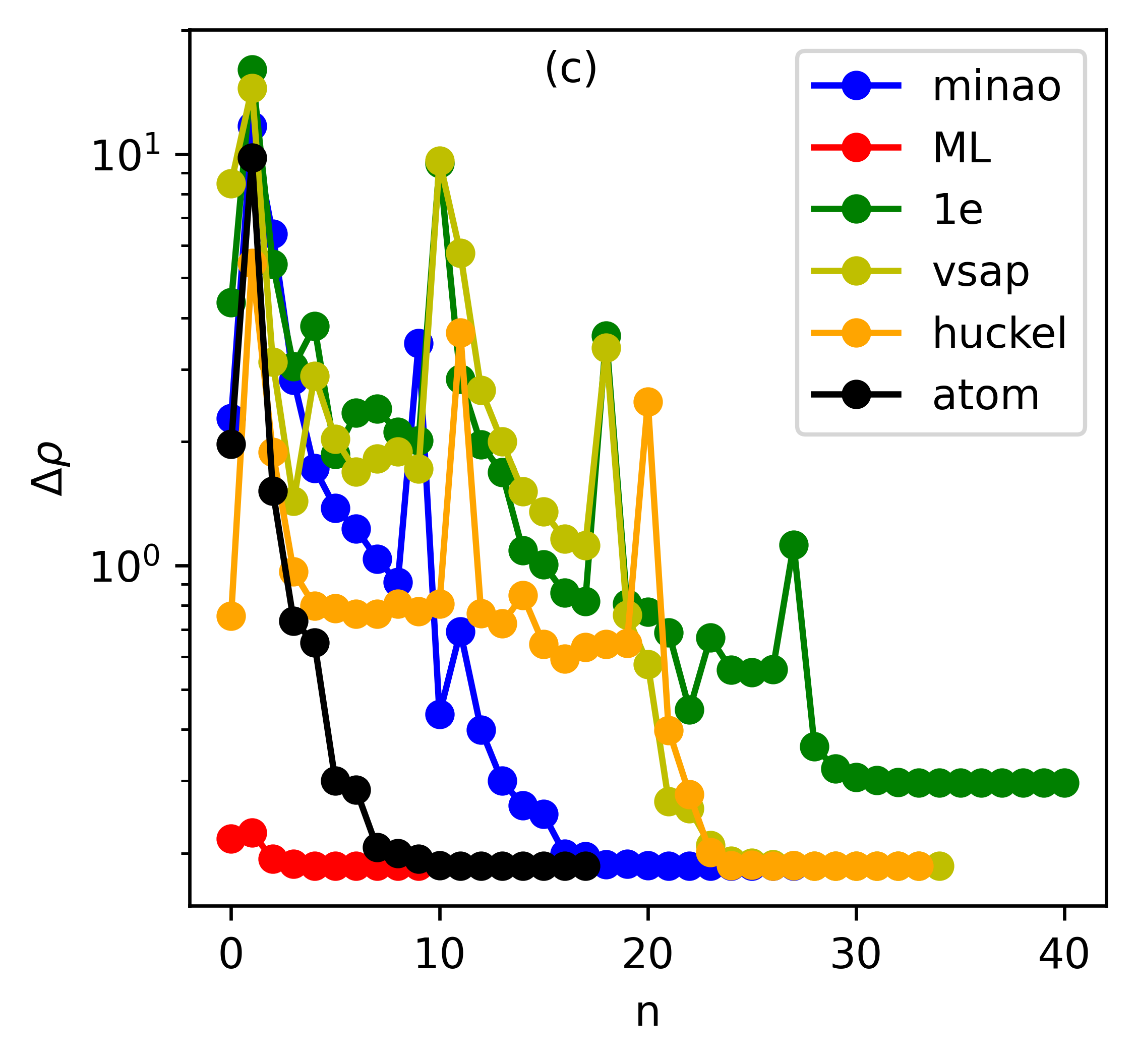}}
    \caption{Analysis of the SCF cycle for \feho. In panels (a) and (b) we show the total
    energy (measured with respect to the ground-state energy) as a function of the iteration number, 
    $n$, for convergence driven by the DIIS and SOS mixing scheme, respectively. In panel (c) we 
    present the norm of the difference between the ground-state converged DM, $\rho^\mathrm{GS}$, 
    and that computed at the $n$-th iteration, $\rho^n$. In this case, we follow the SH: DIIS-driven SCF cycle. 
    For ease of visualization in all plots the $y$ axis is on a logarithmic scale.}  
    \label{fe_ener_den_scf}
    \label{fig:three-graphs} 
\end{figure*}

The path to convergence becomes significantly more oscillatory when one looks at the DIIS
SCF cycle for \feho\ (see Fig.~\ref{fe_ener_den_scf}). This time, the energy and DM 
fluctuations are significantly more pronounced, with the appearance of `spikes' in correspondence 
to some self-consistent steps when using the DIIS solver. These originate from fluctuations
in atomic orbital occupation across different SCF iterations.
Such large fluctuations are suppressed by the SOS mixing scheme, which reinstates
a monotonic approach to the ground-state solution. Most importantly, also for \feho\ 
the ML-constructed DM provides a much more accurate starting point. In fact, it is sufficiently,
accurate that the oscillations are suppressed, regardless of the mixing scheme. Furthermore, 
already at the first iteration energy and DM are extremely accurate. This time, the average 
energy is about $3\times10^{-5}$~Ha away from the converged one (this corresponds to an error 
$1.7\times10^{-6}$\%), while the percentage variation of the DM at the first iteration with respect 
to the ground state is $0.77$\%. 

\subsection{Non-self-consistent forces}
The previous section has shown that the ML DM requires an extremely limited number of 
SCF iterations to achieve the desired convergence and that, even without any 
iteration, it can already provide an accurate estimate of the DFT total energy. Here we 
explore further this second aspect and investigate the accuracy of our ML DM in predicting
a second observable, namely the atomic forces. For this section we consider the \so\ molecule 
in particular, but the results for \ho\ and \feho\ are qualitatively rather similar
{and presented in the SI.}. 
\begin{figure}[!ht]
\includegraphics[width=0.35\textwidth]{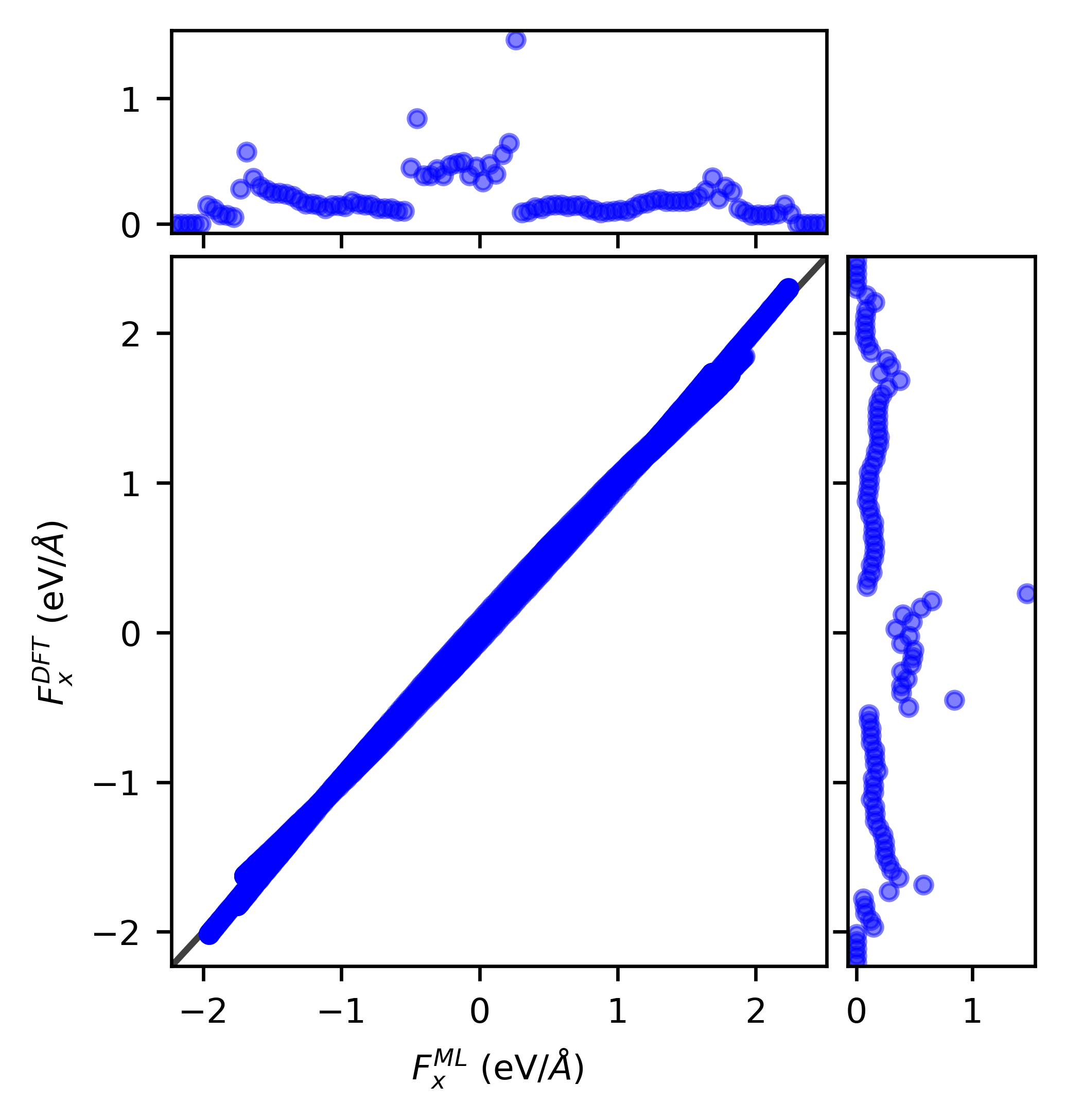}
\vspace{0.1cm}
\includegraphics[width=0.35\textwidth]{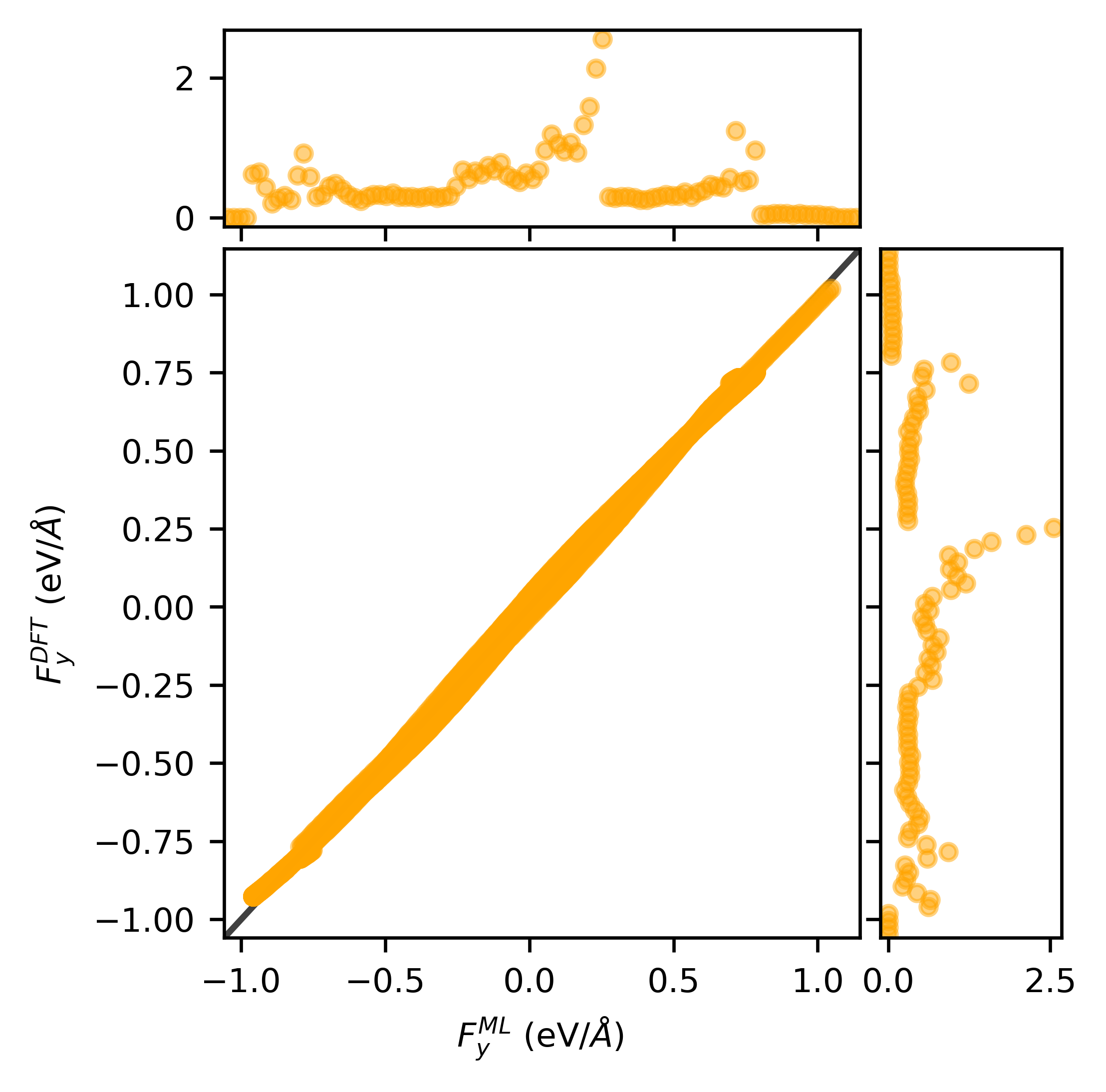}
\caption{Parity plot for the $\alpha=x, y$ component of the atomic forces 
computed by using the ML DM, $F_\alpha^\mathrm{ML}$, with one SCF 
cycle, against the fully converged DFT ones, $F_\alpha^\mathrm{DFT}$.
Data are here presented for a set of 1000 \so\ molecules extracted from the
same molecular dynamics trajectory used to generate the training set. The 
upper panel is for the forces $x$ component, while the lower panel is for the $y$ 
component. The histograms on the side describes the frequency of the forces in 
the test set.}  
\label{parity_forces}
\end{figure}

In figure~\ref{parity_forces} we present the parity plot diagram for the $x$ [upper panel] and $y$ 
[lower panel] component of the atomic forces acting on the atom lying in the $x$-$y$ plane. 
These are computed for a set of 1000 distorted molecules obtained from
the molecular dynamics trajectory used to generate the training set, but never used in
the construction of the neural network. Since, the molecule are, by construction, 
always aligned in the $x$-$y$ plane, there are no forces along $z$. The parity 
plot compares the fully converged DFT forces ($y$ axis) with those predicted from the ML DM 
without operating any SCF iteration ($x$ axis). Points on the parity line are predicted exactly. 
The graphs also show histograms of the distributions of the atomic forces. Note that
along our molecular-dynamics trajectory the forces can be as large as 2~eV/\AA, but typically 
they are concentrated within $\pm$0.25~eV/\AA. 

Clearly, there is an extremely good mapping between the ML-DM-predicted forces and
the exact ones, with the vast majority of the points staying on the parity line. This is
reflected in the almost identical force distributions. Quite interestingly, there is no biased
distribution of errors across the range of force magnitude explored, in contrast to what usually
found for ML force fields, where the largest error is encountered for small forces. The
mean absolute error (MAE) is calculated at 126~meV/\AA\ and 62~meV/\AA, respectively for the 
$x$ and $y$ component. Such error can be further reduced by noticing that the ML-generated 
DM does not necessarily describe an integer number of electrons. This feature can be corrected
by re-scaling the ML-generated DM of a factor $N_e/N_e^\mathrm{ML}$, where $N_e$ is the total 
number of electrons and $N_e^\mathrm{ML}$ that computed using the as-generated ML DM, 
$N_e^\mathrm{ML}=\Tr[\rho^\mathrm{ML}\cdot S]$, with $S$ being the overlap matrix. For 
\so\ we find that typically $N_e^\mathrm{ML}$ is less than 0.1\% different from $N_e$,
but the correction is enough to bring down the MAE to 62~meV/\AA\ and 22~meV/\AA, 
respectively for the $x$ and $y$ component (the parity plots of Fig.~\ref{parity_forces}
have been obtained with the forces computed after such DM re-scaling). This error is significantly 
lower than what typically found for state-of-the-art force fields~\cite{ACE,MTP,JL}. 
Although a thorough comparison is not simple, since the analysis needs to be carried
out with the same molecules, same training set size, etc., this result clearly demonstrates
that predicting the DM to be used in non-self-consistent DFT, can be a valid alternative to
the construction of a force field. Namely, the forces obtained from the ML-predicted DM
can be used as driver for molecular dynamics. This aspects is explored last in the next 
section.

\subsection{Non-self-consistent molecular dynamics}

As a final test, we now perform molecular dynamics simulations by using the 
ML-predicted DM. In particular, we use the forces obtained after rescaling the 
DM by $N_e/N_e^\mathrm{ML}$ and after a single SCF step. Such step is needed, 
since the re-scaled DM seems to have a total energy marginally lower than that of 
the ground state. The molecular dynamics simulations are then performed at 150K 
for \ce{S2O} and \ce{H2O} for total of 0.14 and 0.12 nanoseconds, respectively. In 
both case we use the rotation procedure described in the method section to avoid
the need of an equivariant model. The trajectories obtained are then compared with 
the fully converged {\it ab-initio} ones, and with those computed by performing only 
a single DIIS self-consistent step, starting from the PySCF default `minao' charge 
density.
\begin{figure}[!ht]
\includegraphics[width=0.50\textwidth]{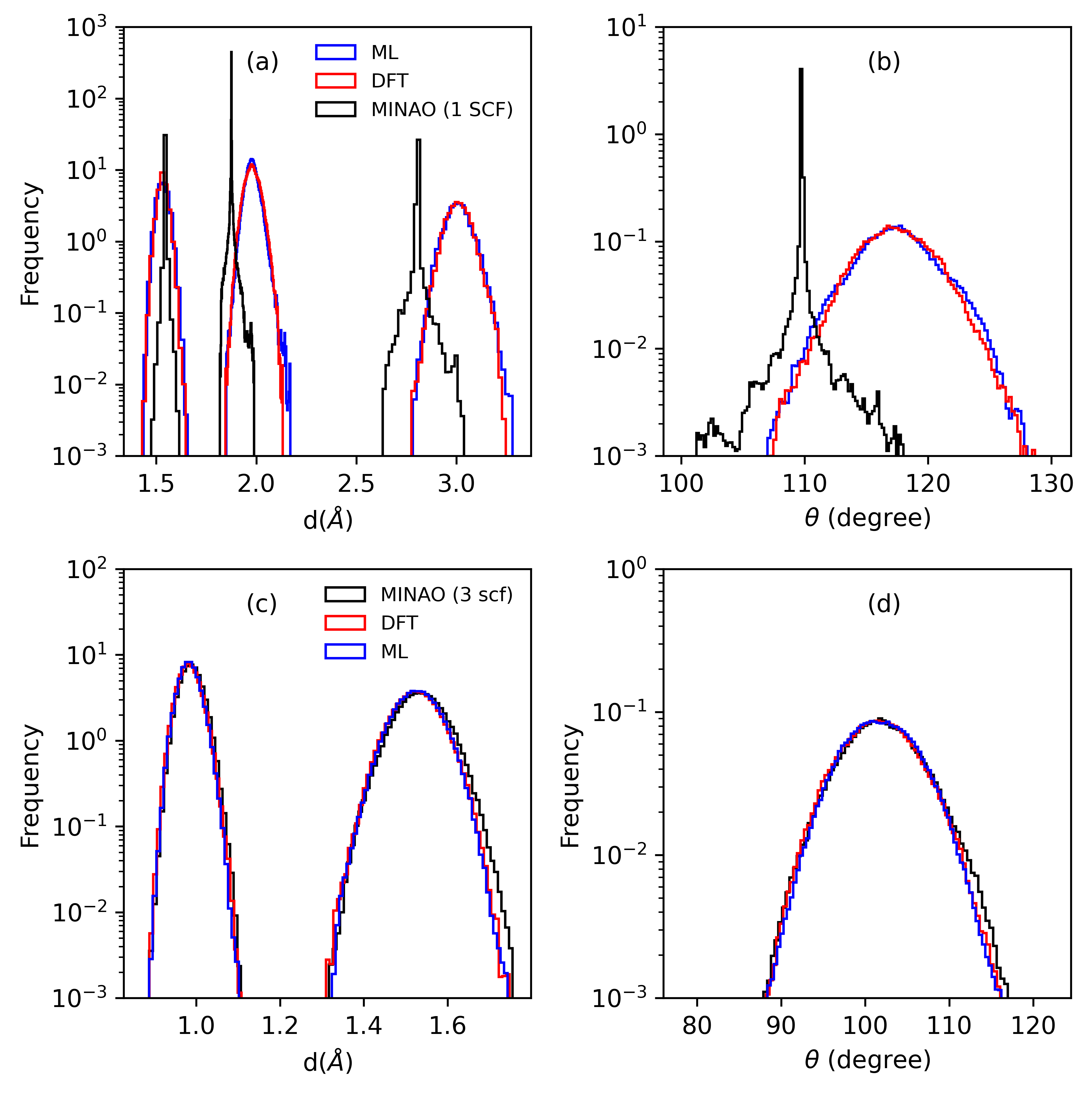}
\caption{Histograms of the bond lengths, $d$, and bond angles, $\theta$, along the molecular dynamics 
trajectories for \so\ [panels (a) and (b)] and \ho\ [panels (c) and (d)]. There are two distinct bond lengths
for \ho, namely O-H and H-H, while there are three for \so, namely two S-O and one S-S.}
\label{bl_ba_s2o}
\end{figure}

The different molecular dynamics trajectories are monitored and compared by looking
at the bond length and bond angle thermal distributions, which are presented in 
Fig.~\ref{bl_ba_s2o} for the two molecules. In the case of \ho\ [panels (c) and (d)] there 
is no significant difference between the various methods, with rather similar distributions. 
This has to be expected considering the speed of convergence of the SCF cycle in this case. 
Thus, we find an average O-H bond length of about 0.9837~\AA\ and an average bond angle of 
101.83$^\circ$, values fully consistent with the static DFT BLYP results, $0.9751$~\AA\ and
$104.14^{\circ}$, and with the experimental values of $0.9578$~\AA\ and $104.47^{\circ}$ \cite{nist}.

The \so\ case is, instead, quite different. From panels (a) and (b) of Fig.~\ref{bl_ba_s2o} one can appreciate
that the ML DM provides an excellent estimate of the fully DFT-converged one, so that the thermal distributions
of bond lengths and angle are rather similar to those obtained with fully {\it ab-initio} molecular dynamics. In 
this case there are three bond lengths corresponding to the S-O bond, the S-S one and the second S-O
distance between the two most peripheral atoms. The centers of the distributions are close to the reported
experimental values of 1.4650~\AA\ (S-O), 1.8834~\AA\ (S-S) and 3.2505~\AA\ (S-S) \cite{nist}, and so is
the bond angle, $117.876^{\circ}$, with the remaining differences being attributed to the choice 
of exchange and correlation energy. This is not the case when the molecular dynamics is performed with
a single SCF step starting from the PySCF `minao' initialization. In fact, the low accuracy in the determination
of the forces results in an average structure presenting a significantly compressed S-S bond and a drastic 
reduction in the bond angle. 

\begin{figure}[!ht]
\includegraphics[width=0.50\textwidth]{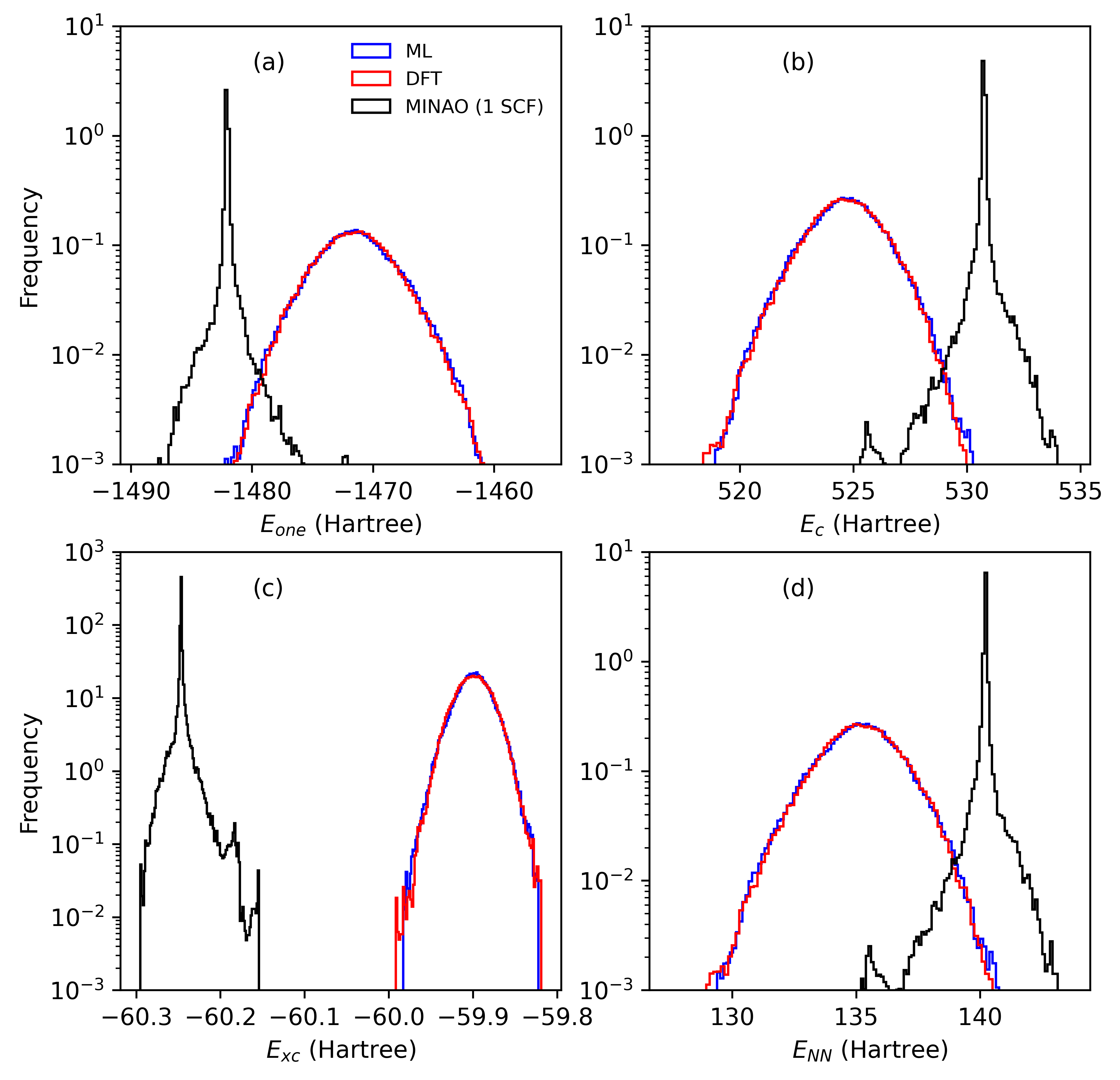}\\
 \caption{\label{ener_com_s2o} Histogram of the various energy components along the different molecular 
 dynamics trajectories for the \ce{S2O} molecule. Here we separate the total energy into one-electron, 
 $H_\mathrm{one}$, Coulomb (Hartree), $H_\mathrm{C}$, exchange-correlation, $H_\mathrm{XC}$, and
 nucleus-nucleus, $H_\mathrm{NN}$, components.}
\end{figure}
Finally, in Fig.~\ref{ener_com_s2o} we present the decomposition of the total energy over the one-electron, 
$H_\mathrm{one}$, Coulomb (Hartree), $H_\mathrm{C}$, exchange-correlation, $H_\mathrm{XC}$, and
nucleus-nucleus, $H_\mathrm{NN}$, components. As expected, since the average structures are erroneously
predicted, the  minao-initialized molecular dynamics provides distributions pretty far from those obtained
with self-consistent DFT. This contrasts the results obtained from our ML DM, which not only describes well
the structure, but also all energy components. 
 
\section{Conclusion}

We have shown that neural networks can be trained to predict the one-particle density matrix required 
by electronic-structure theories constructed over atomic-orbital basis sets. These ML DMs are of sufficient 
quality to be used as initial guess in Kohn-Sham DFT, demonstrating a reduction in the number of the self-consistent
steps needed to achieve convergence over all other common initial choices. Equally important, such DMs
return rather accurate energy and forces even without self-consistency, a fact that enables one to run
inexpensive molecular dynamics simulations, whose computational cost is similar to that needed by 
machine-learning force fields. We have shown here results for three molecules, \ho, \so\ and \feho,
and density functional theory, but the method is agnostic to the system to describe and the choice of
electronic structure theory, as long as this is based on the density matrix. For instance, it can be employed
together with wave-function-based quantum-chemistry methods such as Hartree-Fock. It is also
important to note that, although here the molecule structure is represented by simple Cartesian
coordinates, our proposed method can be implemented with more sophisticated structural descriptors.
These can be constructed equivariant~\cite{Anh} and can be descriptive enough to avoid the need
of using deep-learning machine-learning models~\cite{JL}.

The drawback of our scheme is that the number of matrix elements to predict scales quadratically with the
number of basis functions used in the calculation, so that it becomes increasingly expensive as the system 
size grows. In practice, many of the matrix elements remain rather small and they can be safely neglected
when evaluating the DM for accurate non-self-consistent electronic structure or molecular dynamics. 
{Furthermore, efficiency can be achieved by constructing the ML density matrix over a small basis set and then using it for 
calculations employing larger ones~\cite{Lehtola1}.}

{It is important to note, however, that whether or not a ML strategy for the construction of the DM is favourable against 
other solutions ultimately depends on many considerations. These are characteristic of the system to investigate and the workflow adopted. 
In particular, one may consider three determining factors: 1) the size of the training set needed, namely how many DFT calculations one 
needs to perform for constructing the model); 2) the size of the optimal neural network as the dimension of the DM grows (heavier 
networks may be required as the complexity increases); 3) the workflow in which the method is deployed, namely how many calculations
one has to perform once the ML model has been constructed. All these factors together determine the `computational economy' of any 
ML approach, and a careful assessment must be carried out before a specific computational strategy is selected.}

{In any case, our scheme will become progressively more convenient as the scaling of the overarching electronic 
structure method with the system size becomes more prohibitive.} In this case, the quadratic scaling of the DM construction is
overshadowed by the computational cost to run long self-consistent cycles, and significant savings in 
computational overheads can be achieved. 
This can be the case, for instance, of {non-local exchange-correlation functionals.}

\begin{figure*} [ht!]
    \centerline{\includegraphics[width=0.36\textwidth]{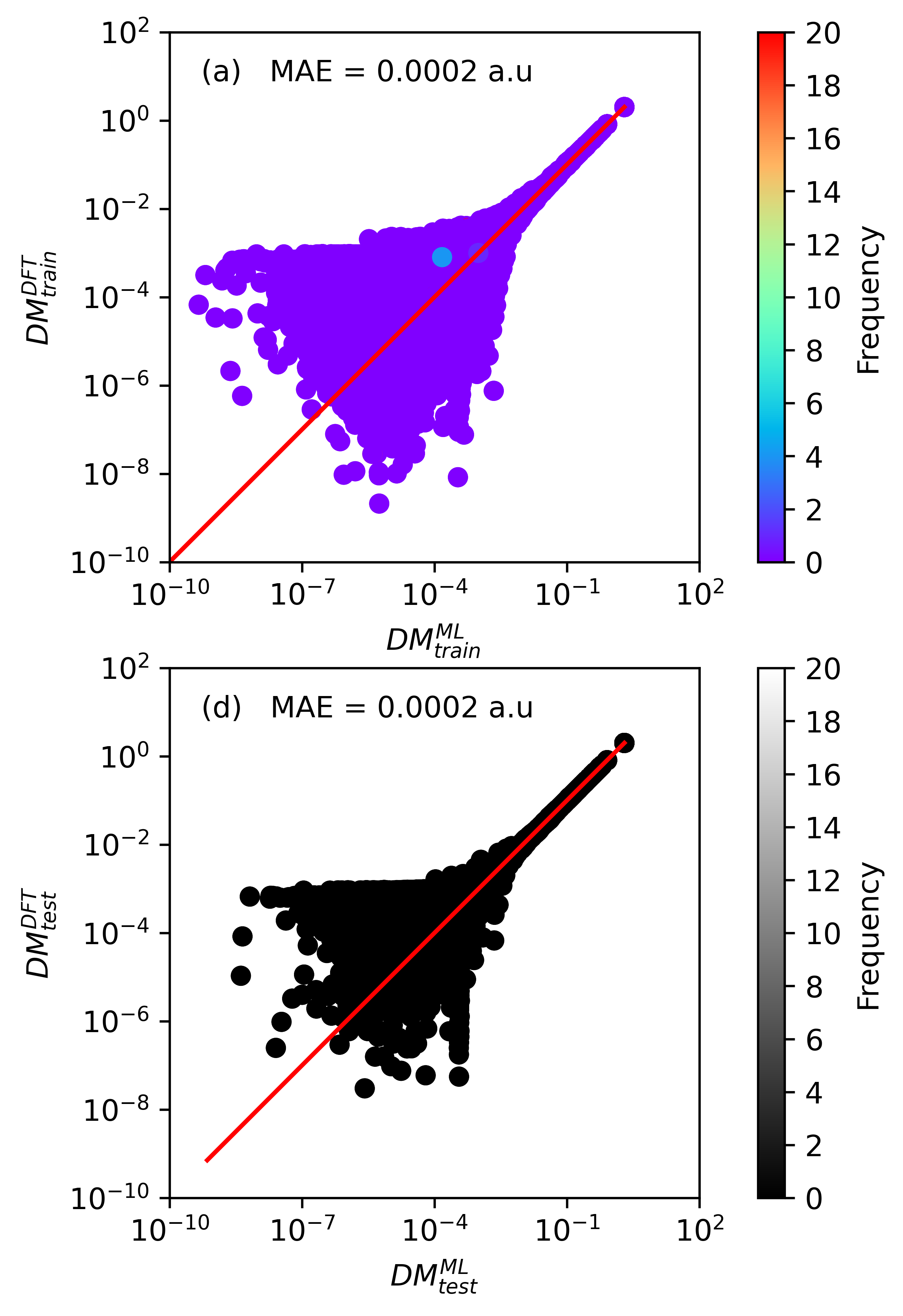}
    \includegraphics[width=0.36\textwidth]{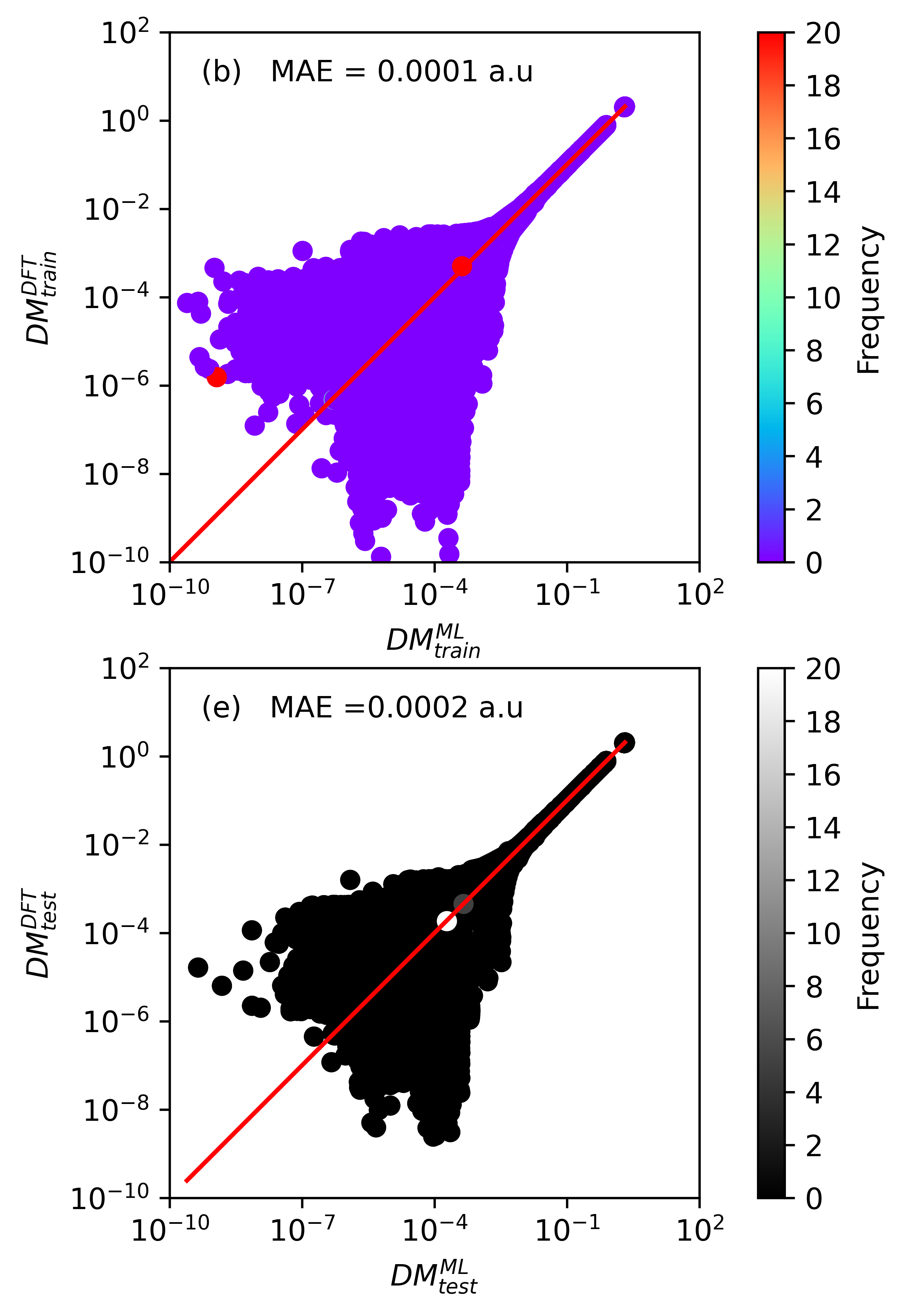}
     \includegraphics[width=0.36\textwidth]{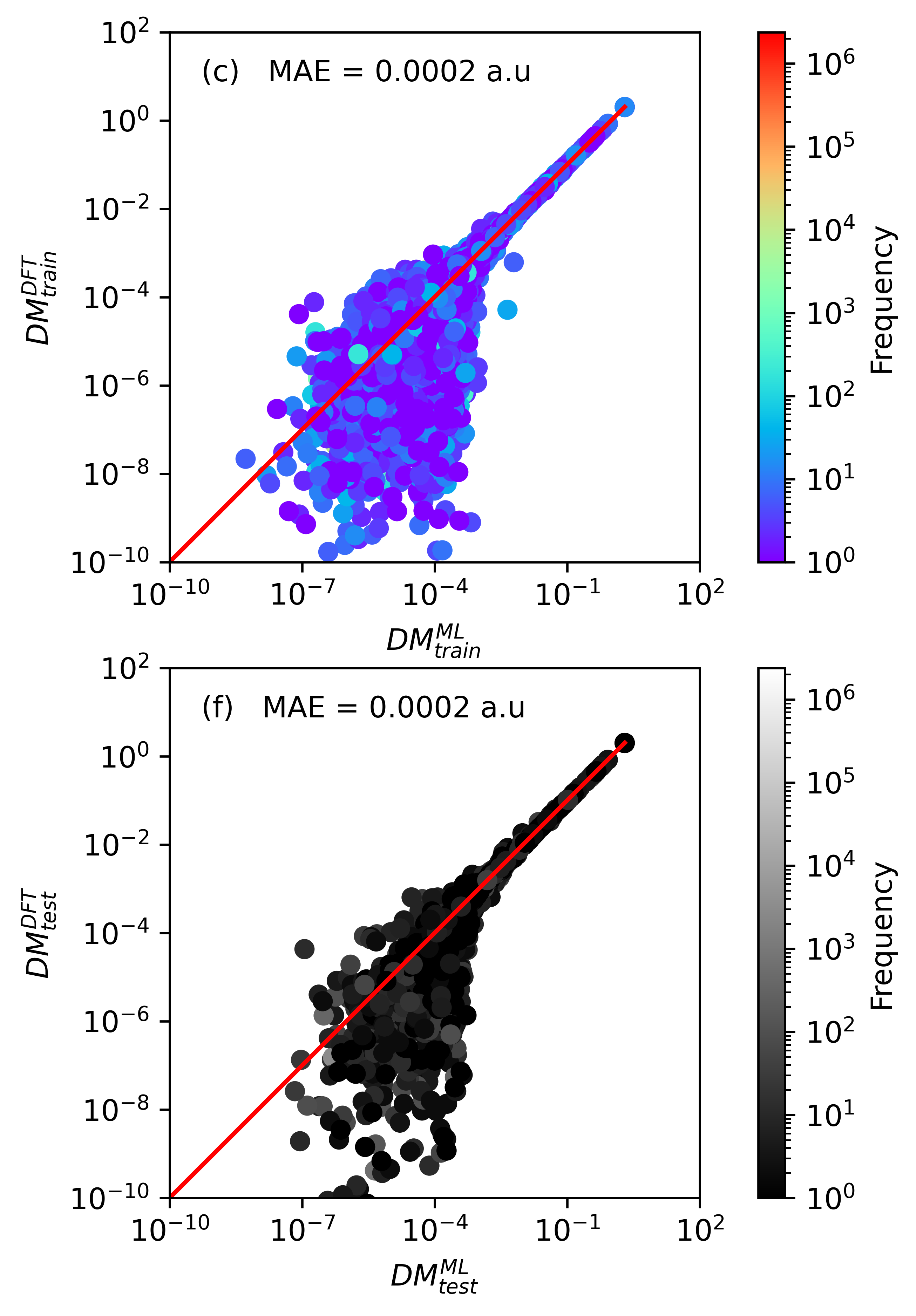}}
    \caption{{Parity plots for \ho\ [panels (a) and (d)], \so\ [panels (b) and (e)], 
    and \feho\ [panels (c) and (f)]. The upper panels are for the training set, and the lower ones 
    for the test one. Each graph reports also the MAE achieved. Note that all the parity plots are in
    logarithmic scale (we plot $|\rho_{ij}|$) and that deviations are only found for the smaller matrix elements.
    The colour code describes the density of given DM matrix-element values.}}  
    \label{h2o_s2o_fe}
    \label{fig:parity} 
\end{figure*}

\begin{acknowledgments}
This work has been supported by the Government of India, NOS Award (K-11015/65/2020-SCD-V/NOS),
and by the Irish Research Council postgraduate program (MC). UP thanks the Qatar National Research Fund
(NPRP12S-0209-190063) for financial support. We acknowledge the DJEI/DES/SFI/HEA Irish Trinity Centre 
for High Performance Computing (TCHPC) for the provision of computational resources.
\end{acknowledgments}

\section{Appendix}
\subsection{Neural-network parity plots}
In Fig.~\ref{h2o_s2o_fe} we present the parity plots (in logarithmic scale) for the elements of the density matrix of 
\ho\ [panels (a) and (d)], \so\ [panels (b) and (e)] and \feho\ [panels (c) and (f)]. The upper panels are for the training set and 
the lower ones for the test set, and the value of the MAE is reported in all cases. As we can see there is an 
excellent agreement between the ML-computed DM and the fully converged DFT one, with most of the points
lying on the parity line. This is true for matrix elements larger than $\sim10^{-2}$, namely for those that have a 
dominant behaviour on all the observable of relevance (e.g. total energy and forces). Larger relative errors are
found for smaller elements, for which the DFT data are already noisy and the ML model can hardly train.
This distribution of errors reflects the rather low MAE reported for all cases (see also Table~\ref{table1}). 
The only exception is for a few points in the training set of \feho, for which some more pronounced deviations are reported. 
These, however, correspond to situations where the DFT calculations struggle to converge (see main test) and the 
DM is the one obtained at the maximum number of self-consistent iterations allowed (hence, not the converged
one). These training points are thus discarded. The test set, instead, contains only examples where full convergence 
is obtained. 

\section{Supporting Information}
{A Supporting Information file is associated to this paper. This includes the convergence analysis for
\so\ and the non-self-consistent energy and forces for \ho\ and \feho. This information is available free of charge via 
the Internet at http://pubs.acs.org}


\begin{thebibliography}{hbp}
\bibitem{theorem} 
Hohenberg P. and Kohn W., Inhomogeneous electron gas, {\it Phys. Rev.} {\bf 1964}, {\it 136}, B864-B871.  

\bibitem{exchange} 
Kohn W. and Sham L.J., Self-consistent equations including exchange and correlation effects, {\it Phys. Rev.} {\bf 1965}, {\it 140}, A1133-1138.

\bibitem{oxford} 
R.G. Parr and W. Yang, {\it Density-Functional Theory of Atoms
and Molecules}, Oxford University Press, New York (1994)  

\bibitem{JLadder}
Perdew J.P. and Schmidt K., Jacob's ladder of density functional approximations for the exchange-correlation energy, {\it AIP Conf. Proc.} {\bf 2001}, {\it 577}, 1-20. 

\bibitem{VASP}Kresse G. and Hafner J., Ab initio molecular dynamics for liquid metals, {\it Phys. Rev. B} {\bf 1993}, {\it 47}, R558-R561.

\bibitem{QE}Giannozzi P. et al., Advanced capabilities for materials modelling with Quantum ESPRESSO, {\it J. Phys.: Condens. Matter} {\bf 2017}, {\it 29}, 465901.

\bibitem{WIEN2K}Blaha P., Schwarz K., Tran F., Laskowski R., Madsen G.K.H. and Marks L.D., WIEN2k: An APW+lo program for calculating the properties of solids, {\it J. Chem. Phys.} {\bf 2020}, {\it 152}, 074101.

\bibitem{Abinit}Romero A.H. et al., ABINIT: Overview and focus on selected capabilities, {\it J. Chem. Phys.} {\bf 2020}, {\it 152}, 124102.

\bibitem{FHIAIMS}Blum V., Gehrke R., Hanke F., Havu P., Havu V., Ren X., Reuter K. and Scheffler M., Ab initio molecular simulations with numeric atom-centered orbitals, {\it Comput. Phys. Commun.} {\bf 2009}, {\it 180}, 2175-2196.

\bibitem{Siesta}Garc\'ia A. et al., {\sc Siesta}: Recent developments and applications, {\it J. Chem. Phys.} {\bf 2020}, {\it 152}, 204108.

\bibitem{PySCF}Sun Q. et al., Recent developments in the PySCF program package, {\it J. Chem. Phys.} {\bf 2020}, {\it 152}, 024109.

\bibitem{rigor}Lejaeghere K. at al., Reproducibility in density functional theory calculations of solids, {\it Science} {\bf 2016}, {\it 351}, aad3000.

\bibitem{Lehtola2023}Lehtola S. and Marques M.A.L., Reproducibility of density functional approximations: How new functionals should be reported, {\it Chem. Phys.} {\bf 2023}, {\it 159}, 114116.

\bibitem{condensed}Wang Y.A. and E.A.~Carter, Orbital-Free Kinetic-Energy Density Functional Theory, in Schwartz S.D. (ed.), Theoretical Methods in Condensed Phase Chemistry. Progress in Theoretical Chemistry and Physics, vol 5. Springer, Dordrecht (2002).

\bibitem{orbital}Chen H. and Zhou A., Orbital-free density functional theory for molecular structure calculations, {\it Numer. Math. Theor. Meth. Appl.} {\bf 2008}, {\it  1}, 1-28. 

\bibitem{recent}Wesolowski T.A. and Wang Y.A., Recent Progress in Orbital-Free Density Functional Theory, World Scientific (Singapore) (2013).  

\bibitem{KieronKE}Li L., Snyder J.C., Pelaschier I.M., Huang J., Niranjan, U.-N., Duncan P., Rupp M., M\"uller K.-R. and Burke, Understanding Machine-Learned Density Functionals, {\it Int. J. Quantum Chem.} {\bf 2016}, {\it 116}, 819-833.

\bibitem{QuantumChem}Szabo A. and Ostlund N.S., Modern Quantum Chemistry: Introduction to Advanced Electronic Structure Theory, Dover Publications, New York (1996).

\bibitem{diis1}Pulay P., Convergence acceleration of iterative sequences - the case of SCF iteration, {\it Chem. Phys. Lett.} {\bf 1980}, {\it 73}, 393-398.

\bibitem{diis2}Pulay P., Improved SCF convergence acceleration, {\it J. Comp. Chem.} {\bf 1982}, {\it 3}, 556-560.  

\bibitem{blackboxscf}Kudin K.N., Scuseria G.E. and Canc\`es E., A black-box self-consistent field convergence algorithm: One step closer, {\it J. Chem.Phys.} {\bf 2002}, {\it 116}, 8255-8261.

\bibitem{damping}Kudin K.N. and Scuseria G.E., Converging self-consistent field equations in quantum chemistry - recent achievements and remaining challenges, {\it ESAIM: M2AN} {\bf 2007}, {\it 41}, 281-296.

\bibitem{outperformdiis}Canc\'es E. and Bris C.L., Can we outperform the DIIS approach for electronic structure calculations?, {\it Int. J. Quantum Chem.} {\bf 2000}, {\it 79}, 82-90.   
 
\bibitem{levelshift}Saunders V.R. and Hillier I.H., Level-shifting method for converging closed shell Hartree-Fock wave-functions, {\it Int. J. Quant. Chem.} {\bf 1973}, {\it 7}, 699-705.   
 
\bibitem{ediis}Bacskay G.B., A quadratically convergent Hartree-Fock (QC-SCF) method - application to closed shell systems, {\it Chem. Phys.} {\bf 1981}, {\it 61}, 385-404. 

\bibitem{Bhattacharyya}Bhattacharyya S.P., Accelerated convergence in SCF calculations and level shifting technique, {\it Chem. Phys. Lett.} {\bf 1978}, {\it 56}, 395-398.  
 
\bibitem{RoothaanHallenergyfunction}Hu X. and Yang W., Accelerating self-consistent field convergence with the augmented Roothaan-Hall energy function, {\it J. Chem. Phys.} {\bf 2010}, {\it 132}, 054109.  

\bibitem{Coiterative}Sun Q., Co-iterative augmented Hessian method for orbital optimization, {\bf 2016}, {\it arXiv preprint arXiv:},1610.08423. 

\bibitem{SOSCF}Sun Q., Yang J. and Chan G.K.-L., A general second order complete active space self-consistent-field solver for large-scale systems, {\it Chem. Phys. Lett.} {\bf 2017}, {\it 683}, 291-299.

\bibitem{Lehtola1}Lehtola S., Assessment of initial guesses for self-consistent field calculations, Superposition of atomic potentials: simple yet efficient, {\it J. Chem. Theo. Comput.} {\bf 2019}, {\it 15}, 1593-1604.  

\bibitem{BurkeD}Brockherde F., Vogt L., Li L., Tuckerman M.E., Burke K. and M\"uller K.-R., Bypassing the Kohn-Sham equations with machine learning, {\it Nat Commun.} {\bf 2017}, {\it 8}, 872.

\bibitem{Chandrasekaran} Chandrasekaran A., Kamal D., Batra R., Kim C., Chen L. and Ramprasad R., Solving the electronic structure problem with machine learning,
{\it npj Comput. Mater.}, {\bf 2019}, {\it 5}, 22.

\bibitem{Ellis}Ellis J.A., Fiedler L., Popoola G.A., Modine N.A., Stephens J.A., Thompson A.P., Cangi A. and Rajamanickam S., Accelerating finite-temperature Kohn-Sham density functional theory with deep neural networks, {\it Phys. Rev. B} {\bf 2021}, {\it 104}, 035120.

\bibitem{Bruno}Focassio B., Domina M., Patil U., Fazzio A. and Sanvito S., Linear Jacobi-Legendre expansion of the charge density for machine learning-accelerated electronic structure calculations, {\it npj Comp. Mater.} {\bf 2023}, {\it 9}, 87.

\bibitem{Grisafi}Grisafi A., Fabrizio A., Meyer B., Wilkins D.M., Corminboeuf C. and Ceriotti M., Transferable Machine-Learning Model of the Electron Density, {\it ACS Cent. Sci.} {\bf 2019}, {\it 5}, 57-64.

\bibitem{Pavanello}Shao X., Paetow L., Tuckerman M.E. and Pavanello M., Machine Learning Electronic Structure Methods Based On The One-Electron Reduced Density Matrix, {\it Nature Commun.} {\bf 2023}, {\it 14}, 6281.

\bibitem{Schutt2019}Sch\"utt K.T., Gastegger M., Tkatchenko A., M\"uller K.-R. and Maurer R.J., Unifying machine learning and quantum chemistry with a deep neural network for molecular wavefunctions. {\it Nat. Commun.} {\bf 2019}, {\it 10}, 5024.

\bibitem{Unke2021}Unke O., Bogojeski M., Gastegger M., Geiger M., Smidt T. and M\"uller K.-R., 
SE(3)-equivariant prediction of molecular wavefunctions and electronic densities, in {\it Advances in Neural Information Processing Systems 34 (NeurIPS 2021)}.

\bibitem{Zhang2022}Zhang L., Onat B., Dusson G., McSloy A., Anand G., Maurer R.J., Ortner C. and Kermode J.R., Equivariant analytical mapping of first principles Hamiltonians to accurate and transferable materials models, {\it npj Comp. Mater.} {\bf 2022}, {\it 8}, 158.

\bibitem{software}Sun Q., Berkelbach T.C., Blunt N.S., Booth G.H., Guo S., Li Z., Liu J., McClain J.D., Sayfutyarova E.R., Sharma S., Wouters S. and Chan G.K.-L., PySCF: the Python-based simulations of chemistry framework, {\it WIREs Comput. Mol. Sci.} {\bf 2018}, {\it 8}, e1340. 

\bibitem{Dunning1989}Dunning T.H., Jr., Gaussian basis sets for use in correlated molecular calculations. I. The atoms boron through neon and hydrogen, {\it J. Chem. Phys.} {\bf 1989}, {\it 90}, 1007-1023.

\bibitem{B88}Becke A.D., Density-functional exchange-energy approximation with correct asymptotic behavior, {\it Phys. Rev. A} {\bf 1988}, {\it 38}, 3098-3100.

\bibitem{LYP}Lee C., Yang W. and Parr R.G., Development of the Colle-Salvetti correlation-energy formula into a functional of the electron density, {\it Phys. Rev. B} {\bf 1988}, {\it 37}, 785-789.

\bibitem{LibXC}Lehtola S., Steigemann C., Oliveira M.J.T. and Marques M.A.L., Recent developments in {\sc libxc} - A comprehensive library of functionals for density functional theory, {\it Software X} {\bf 2018}, {\it 7}, 1-5.


\bibitem{Almlof1982}Alml\"of J., Faegri K. Jr. and Korsell K., Principles for a direct SCF approach to LCAO-MO ab-initio calculations, {\it J. Comp. Chem.} {\bf 1982}, {\it 3}, 385-399.

\bibitem{Lenthe2006}Van Lenthe J.H., Zwaans R., Van Dam H.J.J. and Guest M.F., Starting SCF calculations by superposition of atomic densities, {\it J. Comp. Chem.} {\bf 2006}, {\it 27}, 926-932.

\bibitem{Lehtola2}Lehtola S., Fully numerical calculations on atoms with fractional occupations and range-separated exchange functionals, {\it Phys. Rev. A} {\bf 2020},{\it 101}, 012516. 


\bibitem{Tess}Thomas N., Smidt T., Kearnes S., Yang L., Li L., Kohlhoff K. and Riley P., Tensor field networks: Rotation- and translation-equivariant neural networks for 3D point clouds, {\bf 2018} {\it arXiv}:1802.08219.

\bibitem{lammps}Plimpton J.S., Fast Parallel Algorithms for Short-Range Molecular Dynamics, {\it Comp. Phys.} {\bf 1995}, {\it 117}, 1-19; 
LAMMPS website:{\it https://www.lammps.org}, 
LAMMPS GitHub repository: https://github.com/lammps/lammps

\bibitem{Andrea}Droghetti A., Alf\`e D. and Sanvito S., Assessment of density functional theory for iron(II) molecules across the spin-crossover transition, {\it J. Chem. Phys.} {\bf 2012}, {\it 137}, 124303.

\bibitem{CIPaper}Domingo A., Carvajal M.A. and de Graaf C., Spin crossover in Fe(II) complexes: An ab initio study of ligand $\sigma$-donation, {\it Int. J. Quantum Chem.} {\bf 2010}, {\it 110}, 331-337.

\bibitem{ACE}Drautz R., Atomic cluster expansion for accurate and transferable interatomic potentials, {\it Phys. Rev. B} {\bf 2019}, {\it 99}, 014104.

\bibitem{MTP}Shapeev A.V., Moment tensor potentials: A class of systematically improvable interatomic potentials, {\it Multiscale Modeling \& Simulation} {\bf 2016}, {\it 14}, 1153-1173.

\bibitem{JL}Domina M., Patil U., Cobelli M. and Sanvito S., Cluster expansion constructed over Jacobi-Legendre polynomials for accurate force fields, {\it Phys. Rev. B} {\bf 2023}, {\it 108}, 094102.

\bibitem{nist} Johnson III R.D. (Ed.), NIST Computational Chemistry Comparisons and Benchmark Database, NIST Standard Reference Database, Number, Release {\bf 18}, NIST, Gaithersburg, MD, 2016. 

\bibitem{Anh}Nguyen V.H.A. and Lunghi A., Predicting tensorial molecular properties with equivariant machine learning models, {\it Phys. Rev. B} {\bf 2022}, {\it 105}, 165131.

\end{thebibliography}
\end{document}